\documentclass[iop,apj]{emulateapj}

\usepackage{natbib}
\usepackage{graphicx}	
\usepackage{amsmath}	
\usepackage{amssymb}	
\usepackage{units}
\usepackage{enumitem}
\usepackage{bm}
\usepackage{color}
\usepackage[breaklinks,colorlinks,citecolor=blue,linkcolor=red]{hyperref} 

\def\msun{{\rm\,M_\odot}}

\def\msun{{\rm\,M_\odot}}

\newcommand{\kms}{\, {\rm km\, s}^{-1}}

\newcommand{\be}{\begin{equation}}
\newcommand{\ee}{\end{equation}}

\def\h2{${\rm\,H_2}$}

\newcommand{\beq}{\begin{equation}}
\newcommand{\beqa}{\begin{eqnarray}}
		 \newcommand{\eeq}{\end{equation}}
\newcommand{\eeqa}{\end{eqnarray}}

\begin{document}

\title{The Challenge to MOND from ultra faint dwarf galaxies}
\author{Mohammadtaher Safarzadeh \& Abraham Loeb}
\affil{Center for Astrophysics | Harvard \& Smithsonian, 60 Garden Street, Cambridge, MA; \\
\href{mailto:mohammadtaher.safarzadeh@cfa.harvard.edu}{mohammadtaher.safarzadeh@cfa.harvard.edu}}

\begin{abstract}
Modified Newtonian Dynamics (MOND) at low acceleration has been astonishingly powerful at explaining the rotation curve of galaxies 
and the relation between the baryonic content of the galaxies and their observed circular velocity, known as the Baryonic Tully-Fisher Relationship (BTFR). 
It is known that MOND fails at explaining the observed velocity dispersion of the ultra-faint dwarf galaxies (UFDs) 
with the justification that UFDs are more prone to tidal disruption in MOND compared to cold dark matter model. We show that: (i) the ratio of tidal to internal acceleration in UFDs is extremely low, (ii) there is no correlation between the deviation of UFDs from MOND's prediction as a function of tidal susceptibility, and (iii) recent constraints from Gaia proper motion analysis on the orbital parameters of the UFDs exacerbates the challenge to MOND. 
In particular, Gaia data indicates that Ursa Major I is experiencing a recent infall into the Milky Way's halo, and its inconsistency with MOND at 7-$\sigma$ level can not be attributed to being an early infall satellite. 
Moreover, the new data from Gaia DR2 shows Willman I to have the least eccentric orbit of all UFDs, and its deviation from MOND at 4-$\sigma$ level can not be attributed to a highly eccentric orbit as previously suggested.
Finally, given that Tuc III is the only UFD observed to show tidal features, Reticulum II and Segue I are two other UFDs that potentially challenge MOND as they have comparable galactocentric distances to Tuc III while showing no tidal features. 
\end{abstract}


\section{Introduction}
\label{sec:intro}
Despite all the efforts from multiple theoretical and experimental directions, the nature of dark matter remains unknown \citep[see][ for a recent review]{Profumo2019}.
One important clue to its nature could come from the fact that it should manifest itself on Galactic scales as is successfully described by Modified Newtonian Dynamics \citep[MOND; ][]{Milgrom1983ApJ}.
MOND has specific predictions for systems with acceleration below $a_0=1.2 \times 10^{-8} \rm cm~s^{-2}=3.8~\rm pc/Myr^2$ such that the acceleration becomes $g=\sqrt{g_N a_0}$, where $g_N$ is the Newton's acceleration.
Intriguingly, the value of $a_0$ is related to the Hubble constant by $a_0\approx c H_0/6$ where $c$ is the speed of light.

Two successes of MOND on galactic scales are: i) explaining the observed steep relation with low scatter between baryonic mass in galaxies and their circular velocity
known as Baryonic Tully-Fisher Relationship \citep[BTRF; ][]{BTFR}, and ii) explaining the rotation curve of spiral galaxies \citep{Rubin1980ApJ}.  and the. While in the standard $\Lambda \rm CDM$ cosmology fine-tuning of baryonic physics is required to recover the BTRF \citep{McGaugh2000ApJ,Abadi2003ApJ,Governato2007MNRAS,Bouch2010ApJ,Vogelsberger2013MNRAS,Chan2015MNRAS,Sales2017MNRAS}, 
MOND can reproduce the observed relationship with no fine-tuning as it requires no dark matter in its formulation \citep{McGaugh2005ApJ}. Likewise, the modified dynamics have been astonishigly successful at reproducing the observed rotation curve of the galaxies \citep{SM02}.

In the limit of isolated systems, MOND's prediction for the 1 D velocity dispersion of a dispersion supported spherical galaxies for $g_{in}\ll a_0$ is given by:
\be
\sigma_{iso}\approx \left(\frac{4}{81} a_0 G M_*\right)^{1/4},
\ee
where $M_*$ is the stellar mass of the galaxy, and the internal acceleration, $g_{in}$,  is given by:
\be
g_{in}\approx \frac{3 \sigma_{iso}^2}{r_{1/2}},
\ee
where $r_{1/2}$ is the half light radius. 
However, if the galaxy is a satellite of the Milky Way in the regime of $g_{in}<g_{ex}<a_0$, where $g_{ex}$ is the gravitational acceleration of the MW, given by:
\be
g_{ex}\approx \frac{V_{MW}^2}{D_{GC}},
\ee
MOND's predicted dispersion due to the external field effect \citep[EFE; ][]{MM2013ApJ} is given by:
\be
\sigma_{efe}\approx \left(\frac{a_0 G M_*}{3 g_{ex} r_{1/2}}\right)^{1/2},
\ee
where $V_{MW}$ is the circular velocity of the MW, for which we adopt a constant value of $200~\kms$ for all the calculations in this work \citep{Reid2014ApJ}. 
$D_{GC}$ is the galactocentric distance of the satellite from the MW. We note that the EFE of M31 on MW's satellites is irrelevant since the internal acceleration of the satellites are stronger than M31's external acceleration.

By definition, satellites with luminosity less than $10^5 L_{\odot}$ are considered to be ultra-faint dwarf galaxies (UFDs).
In the standard $\rm \Lambda CDM$ cosmology, UFDs are dark matter-dominated systems with quenched star formation history \citep{SG2007ApJ,Simon2019ARAA}. 
New UFDs were recently discovered in the Dark Energy Survey \citep{Bechtol2015ApJ,Koposov2015ApJ}. As relics from the early universe, UFDs are testbeds for examining various theories from the standard collisionless dark matter \citep{Nadler2021PhRvL}, to fuzzy dark matter \citep{SS2020ApJ,Burkert2020ApJ,Hayashi2021} and MOND \citep{MW2010ApJ}.

A notable example of the success of MOND for the velocity dispersion of the Milky Way's satellites is MOND's prediction for the ultra diffuse galaxy Crater 2 \citep{McGaugh2016ApJ} which has been proved by the observation \citep{Caldwell2017ApJ}. 
This is an important prediction for MOND since Crater 2 is free of current tidal interference (though it may have suffered in the past).
However, it has been shown that MOND's prediction for the velocity dispersion of the UFDs does not match the observed values, but the discrepancy has been attributed to tidal susceptibility \citep{MW2010ApJ}. 
In this work, we revisit this claim in light of the recently discovered UFDs. We argue that tidal susceptibility does not seem to justify the deviation of MOND's predictions from the observed velocity dispersion of UFDs. 

The structure of this paper is as follows: In \S2 we present our method of approximating tidal susceptibility. We investigate which UFDs are within the regime that MOND's EFE.
We point out a few UFDs that challenge MOND and present the caveats around those systems. In \S3, we summarize our results and discuss the most promising avenues towards understanding the nature of dark matter.

\section{DATA}\label{sec:data}
The data on the velocity dispersion and half-light radius of the UFDs are primarily collected from the compilation of \citet{Wolf2010}.
We use the estimated galactocentric distances of the UFDs from the recent proper motion analysis of the Gaia DR2 \citep{Fritz2018,Simon2018ApJGaia}.
The stellar mass of the UFDs is from the compilation of \citet{McConnachie2012}. For the UFDs that are not presented in \citet{Wolf2010} we have collected the data from the following references:
Grus II and Tucana IV from \citet{Simon2020ApJ}, Eridanus II from \citet{Li2017ApJ} and its stellar mass from \citet{Gallart2021ApJ} and \citet{Bechtol2015ApJ}. 
We obtain the data for Horologium I (Hor I) from \citet{Koposov2015ApJRetIIHolI} and its stellar mass from \citet{Bechtol2015ApJ} which is consistent with the data reported in \citet{Jerjen2018}.
Hor I is close to the Magellanic Clouds, and whether this is a satellite of the Large Magellanic Clouds (LMC) is discussed in \citet{Nagasawa2018ApJ}, and we elaborate on in the next section. The data for Triangulum II is obtained from \citet{Kirby2017ApJ}, and for Segue 2 from \citet{Kirby2013ApJSeg2}. The data for Reticulum II is obtained from \citet{Simon2015ApJRetII}. For Bootes I we adopt a velocity dispersion of 2.4 $\kms$ which comes from the majority cold component of stellar population \citep{Koposov2011}. 
As Leo T is a gas-rich satellite with total HI mass of $4\times10^5~\msun$ \citep{Adams2018AA} we consider a total baryonic mass of $\log(M_b/M_{\odot})=5.75$ for this system.
In cases where there is no estimate of the stellar mass of the system, we use the relationship between the metallicity of the system and its stellar mass derived from dwarf satellites of the MW \citep{Kirby2013ApJ}.

\section{Method}\label{sec:method}

\begin{figure}
\hspace{-0.2in}
\centering
\includegraphics[width=\columnwidth]{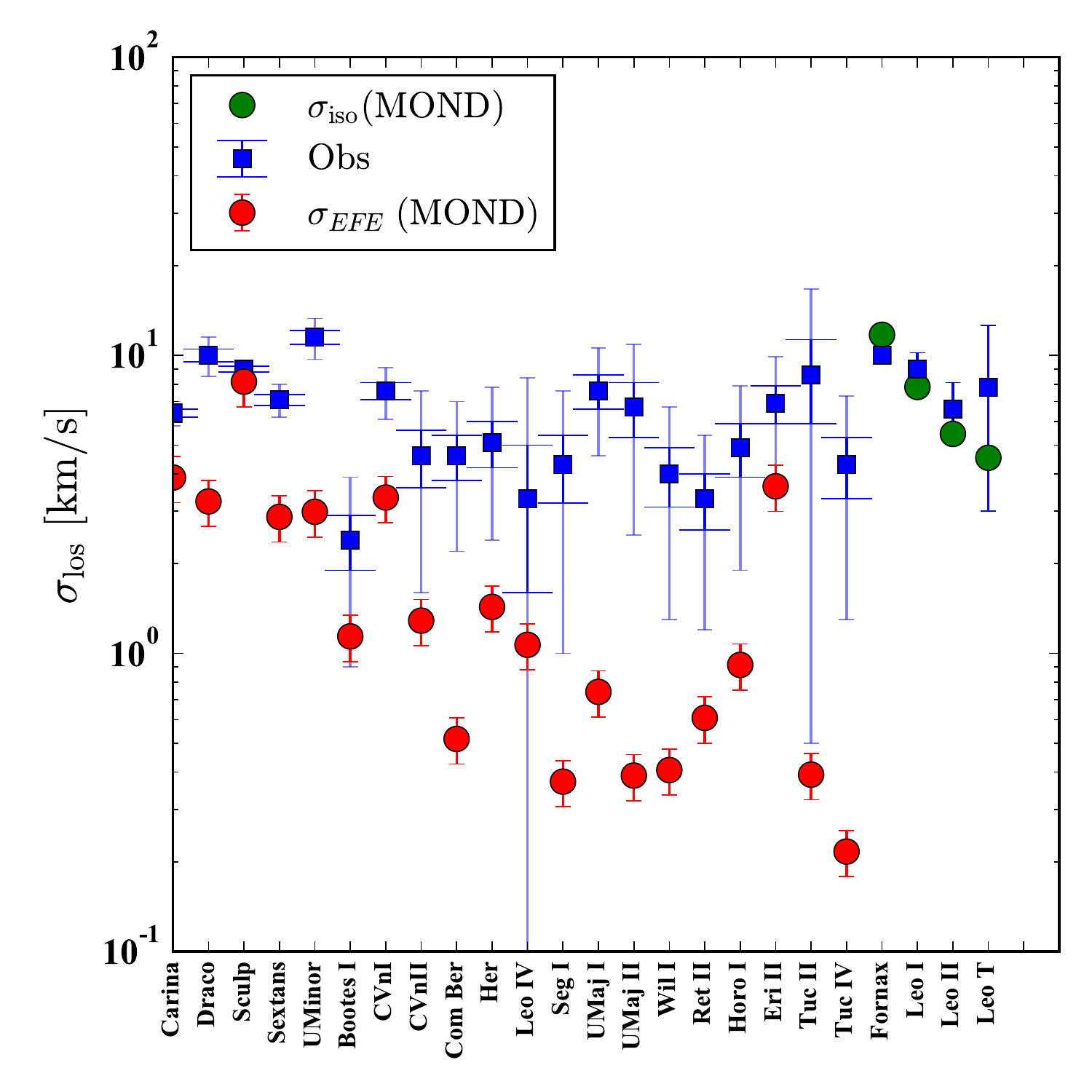}
\caption{The predicted velocity dispersion of the UFDs and dwarf spheroidal galaxies from MOND are shown in red circles. The observed line of sight values (blue squares) of velocity dispersion of the UFDs and dwarf spheroidal galaxies are shown as blue squares with the longer cap error bars are showing 1-$\sigma$ (1 std) error, and the longer and thinner caps showing the 3-$\sigma$ error. We have divided the sample into two groups: those with $g_{in}/g_{ex}<1$ shown with red points, and those with $g_{in}>g_{ext}$ shown with green points. For the former category, we use the $\sigma_{efe}$ of MOND, and for the latter systems, we use the isolated estimate prediction, $\sigma_{iso}$, of MOND. The errorbars for MOND predictions come from assuming the $V_{\rm MW}$ a range between 170 $\kms$ to 230 $\kms$.}
\label{fig_1}
\end{figure}

To begin with, we display the predicted velocity dispersion of MOND for the UFDs as compared to the observed line of sight estimates in Figure \ref{fig_1}.
We have divided the satellites into two groups, one in which the $g_{in}<g_{ex}$ as shown in red points, and those that internal acceleration is larger than $g_{ex}$ as shown in green points.
The squares show the observational estimates, the longer cap error bars are showing the 1-$\sigma$ (1 standard deviation) error, and the longer and thinner caps showing the 3-$\sigma$ error bars.

To investigate whether tidal susceptibility can explain the deviation of MOND's prediction and the observed velocity dispersion of the UFDs, we calculate the ratio of the tidal acceleration and the internal acceleration of a satellite. 
The tidal accelerations are estimated by:
\be
g_t=g_{ex}\frac{2 r_{1/2}}{D_{GC}}.
\ee

\begin{figure}
\hspace{-0.2in}
\centering
\includegraphics[width=\columnwidth]{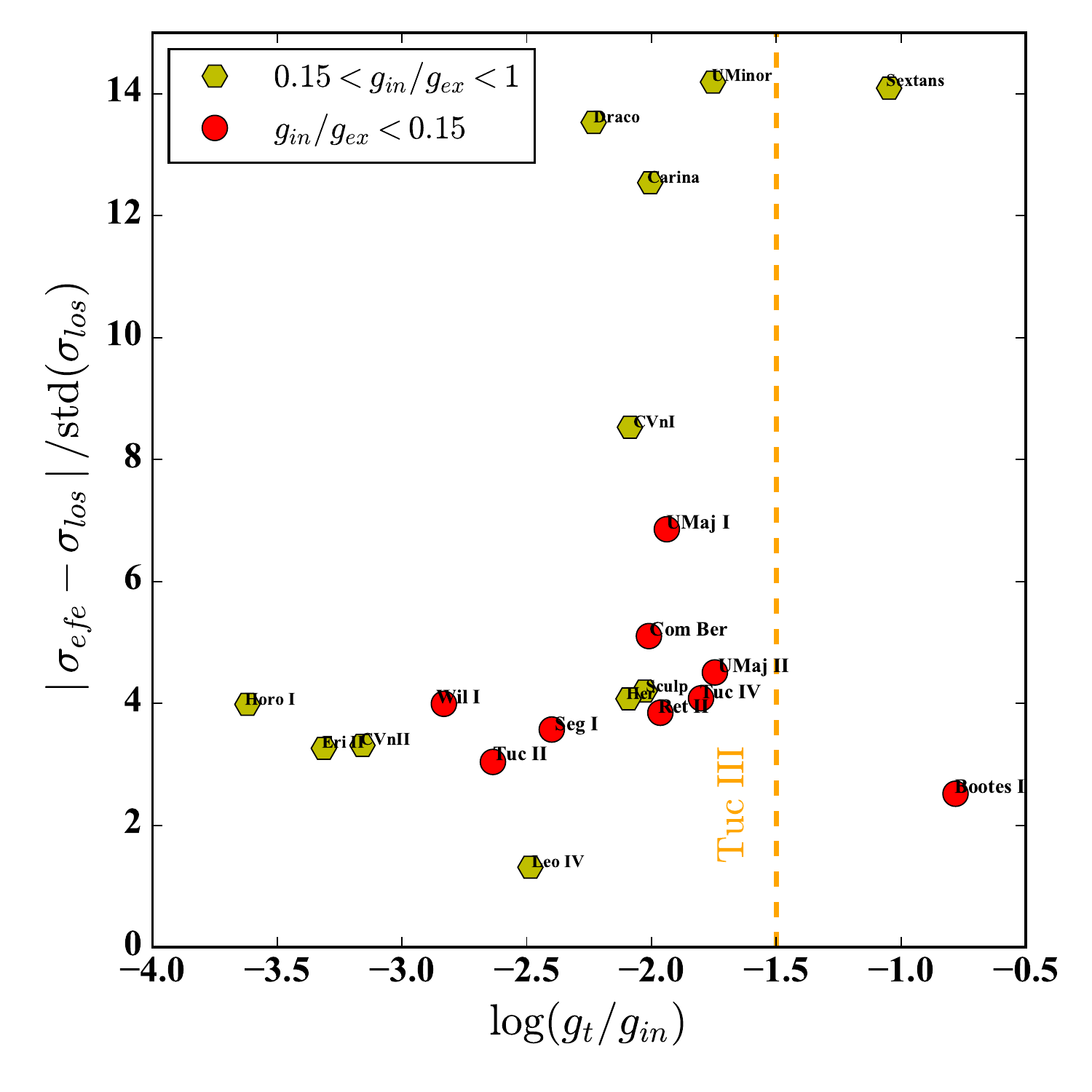}
\caption{Deviation of the predicted MOND velocity dispersion of the UFDs and their observed value as a function of the $\log(g_t/g_{in})$ for systems in which $g_{in}<g_{ex}<a_0$. 
\emph{Yellow hexagons:} for systems with $g_{in}<g_{ex}<1$. \emph{Red points:} for systems with $g_{in}/g_{ex}<0.15$. 
Tuc III is shown with a vertical line since, at the moment, only an upper limit exists for its velocity dispersion, $\sigma_{los}<1.5\kms$ at 95\% confidence. Therefore, the true ratio of tidal over the internal acceleration of Tuc III could be any value smaller than $10^{-1.5}$. }
\label{fig_2}
\end{figure}

The yellow hexagons in Figure \ref{fig_2} show the satellites where $0.15\le g_{in}/g_{ex} \le 1$, and in the red points show the system where $g_{in}/g_{ex}\le 0.15$.
The reason for this separation is that the EFE prediction is only valid when $g_{in}\ll g_{ex}$, and its validity when $g_{in}\approx g_{ex}$ is unkonwn \citep{MM2013ApJ}.
Therefore, we show the results once for all the sample with $g_{in}/g_{ex}<1$, and another time for when $g_{in}/g_{ex}<0.15$, a regime in which EFE is shown to give consistent results with the observations.
The value of 0.15 is chosen based on \citet{McGaugh2016ApJ} EFE model's success in predicting the correct velocity dispersion for Crater II, which has a ratio of $g_{in}/g_{ex}\approx0.12$.

The red points in Figure \ref{fig_2} imply that if tidal susceptibility is the reason for the discrepancy between the predictions of MOND and the observed $\sigma_{los}$ of the UFDs, we have to see a positive correlation between the ratio of the tidal over internal acceleration and the deviation of MOND from the observed velocity dispersion. Does such a correlation exist? With low number statistics, one can either argue for or against a positive correlation. 
However, the fact that Willman I still deviates from the MOND's prediction at four $\sigma$ remains a challenge to MOND. Moreover, Hor I with $g_{in}/g_{ex}\approx0.3$ stands at 4 $\sigma$ from the EFE prediction while having the lowest of the $g_t/g_{in}$ in our sample.

\begin{figure}
\hspace{-0.2in}
\centering
\includegraphics[width=\columnwidth]{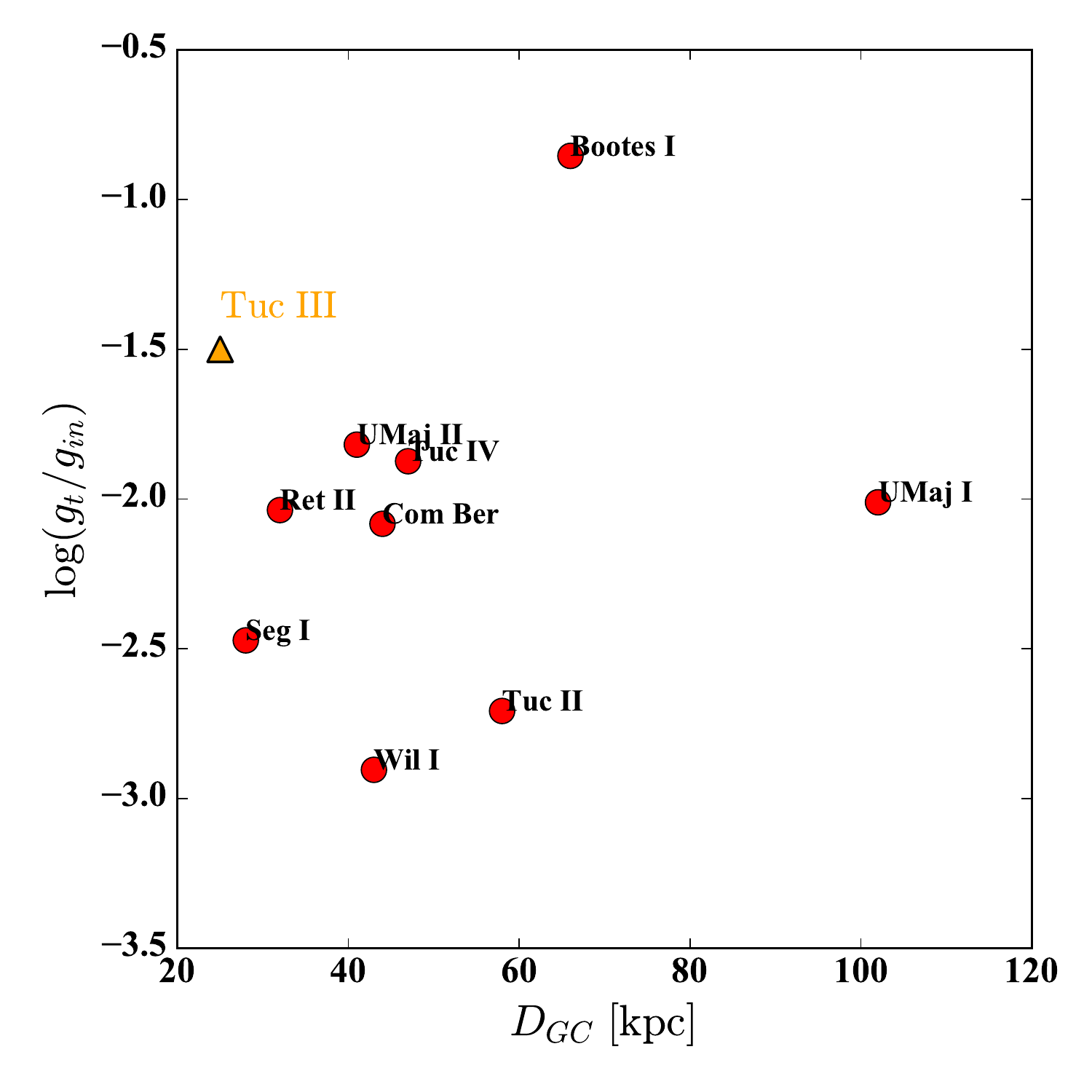}
\caption{The distribution of MW satellites with $g_{in}/g_{ex}<0.15$ in galactocentric distance vs. the ratio of tidal to internal acceleration. Tuc III is shown with a triangle since we only know the lower limit on $g_t/g_{in}$ for Tuc III. The fact that tidal features are observed for Tuc III but not for Ret II or Seg I is a concerning fact. However, this could be justified given the highly eccentric orbit of Tuc III with a pericenter approach of $3^{+1}_{-1}$ kpc. }
\label{fig_3}
\end{figure}

But the most stringent constraint comes from Ret II due to its proximity to the Sun for the following reason:
Tucana III (Tuc III) is the first UFD observed to show a velocity gradient of $8.0\pm0.4 \kms \rm deg^{-1}$, suggesting the galaxy is ongoing tidal disruption \citep{Li2018ApJ}.
\citet{Simon2017ApJTucIII} estimated the $\sigma_{los}$ to be less than 1.5 $\kms$ at 95\% C.I. 
With a half-light radius of about 34 pc and a distance of about 25 kpc, Tuc III is our nearest neighbor UFD with $g_{in}/g_{ex}\approx 0.04$.
Given the old stellar population of the UFDs, one can adopt a value for the stellar mass-to-light ratio of $M_*/L_V\approx 2 M_{\odot}/L_{\odot}$ assuming standard initial mass function \citep{Simon2019ARAA}. 
This leads to an estimate of the stellar mass of Tuc III to be $M_*\approx1500 \msun$.
The same estimate yields a stellar mass of about 5,000 $\msun$ for Ret II. With $\sigma_{los}=1.5\kms$ for Tuc III (which is the estimated lower limit for this system), the ratio of the tidal and internal acceleration for Tuc III is $\log(g_t/g_{in})=-1.5$.
We show this with a vertical line in the bottom panel of Figure \ref{fig_2}. This value is comparable to the ratio of tidal over internal acceleration for Ret II. 

Ret II and Tuc III have similar Galactocentric distances, implying that if we see tidal features in Tuc III, we would also expect to see tidal features in Ret II as well, but no such features have been observed.
However, one can argue that the orbital motion of the two satellites is different: 
based on Gaia DR2, \citet{Simon2018ApJGaia} estimates a pericenter and apocenter distance for Tuc III to be $3^{+1}_{-1}$ kpc and $49^{+6}_{-5}$ kpc with an orbital period of 0.7 Gyr,
and the corresponding values for Ret II are derived from being $29^{+14}_{-6}$ kpc for pericenter approach and $91^{+91}_{-39}$ kpc for apocenter approach with an orbital period of 1.6 Gyr. These numbers agree with the results reported in \citet{Fritz2018}.
The inferred orbital motion based on Gaia DR2 suggests that despite their current similar galactocentric distance, Tuc III has been more prone to tidal disruption as opposed to Ret II. 
However, these orbital motion estimates are based on cold dark matter potential for the MW and whether the same orbital dynamics would be inferred in MOND remains to be verified with detailed orbital motion simulation.
Moreover, we note that although Ret II is an elongated UFD with an ellipticity of about $\epsilon=0.6$, there have been no signs of tidal features within our field of view \citep{MutluPakdil2018ApJ}.

Figure \ref{fig_3} shows the distribution of the satellites in the galactocentric distance and $\log(g_t/g_i)$ plane. Other than Ret II, Seg I also has a comparable galactocentric distance to Tuc III. 
The pericenter and apocenter approach for Segue I (Seg I) is estimated to be $20^{+4}_{-5}$ kpc and $61^{+34}_{-18}$ kpc with an orbital period of about 1.1 Gyr. This makes Seg I more similar to Tuc III than Ret II. 
However, \citet{Simon2011ApJSegI} does not find evidence suggesting Seg I being tidally disrupted. 

Willman I stands 4 $\sigma$ away from the MOND's prediction. This satellite has a pericenter and apocenter approach of $44^{+15}_{-19}$ and $53^{+57}_{-13}$ kpc respectively, which yields this UFD the lowest orbital eccentricity of $e\approx0.2$. While the deviation of Willman I was identified in \citet{MW2010ApJ}, this was attributed to speculating a highly eccentric orbit for Willman I while the new data from Gaia DR2 actually suggests Willman I to be actually on the least eccentric orbit of all the UFDs \citep{Simon2018ApJGaia}.
 
Similar to Willman I, Hor I also stands 4 $\sigma$. away from MOND's prediction; however, with the ratio of $g_{in}/g_{ex}\approx0.3$, it is not clear whether the isolated or the EFE regime is adequate for predicting its internal dynamics. 
Furthermore, the close proximity of Hor I to the LMC \citep{Nagasawa2018ApJ} might further complicate the expected dynamics of this satellite, although $g_{in}/g_{ex}$ with respect to the Magellanic clouds remains less than 1 for this satellite.

Perhaps the most deviant of all the satellites is U Maj I standing about 7 $\sigma$ from MOND's prediction. This deviation was attributed to the early infall time of this satellite in \citep{MW2010ApJ}. 
While early analysis of the infall times of the satellites categorized this satellite as indeed one with an early infall time of about 8 Gyr However \citep{Rocha2012MNRAS}, recent Gaia DR2 suggests that this satellite is actually entering the MW's halo recently in the past $1.5^{+5.1}_{-1.6}$ Gyr \citep{Fillingham2019}.

\begin{figure}
\hspace{-0.2in}
\centering
\includegraphics[width=\columnwidth]{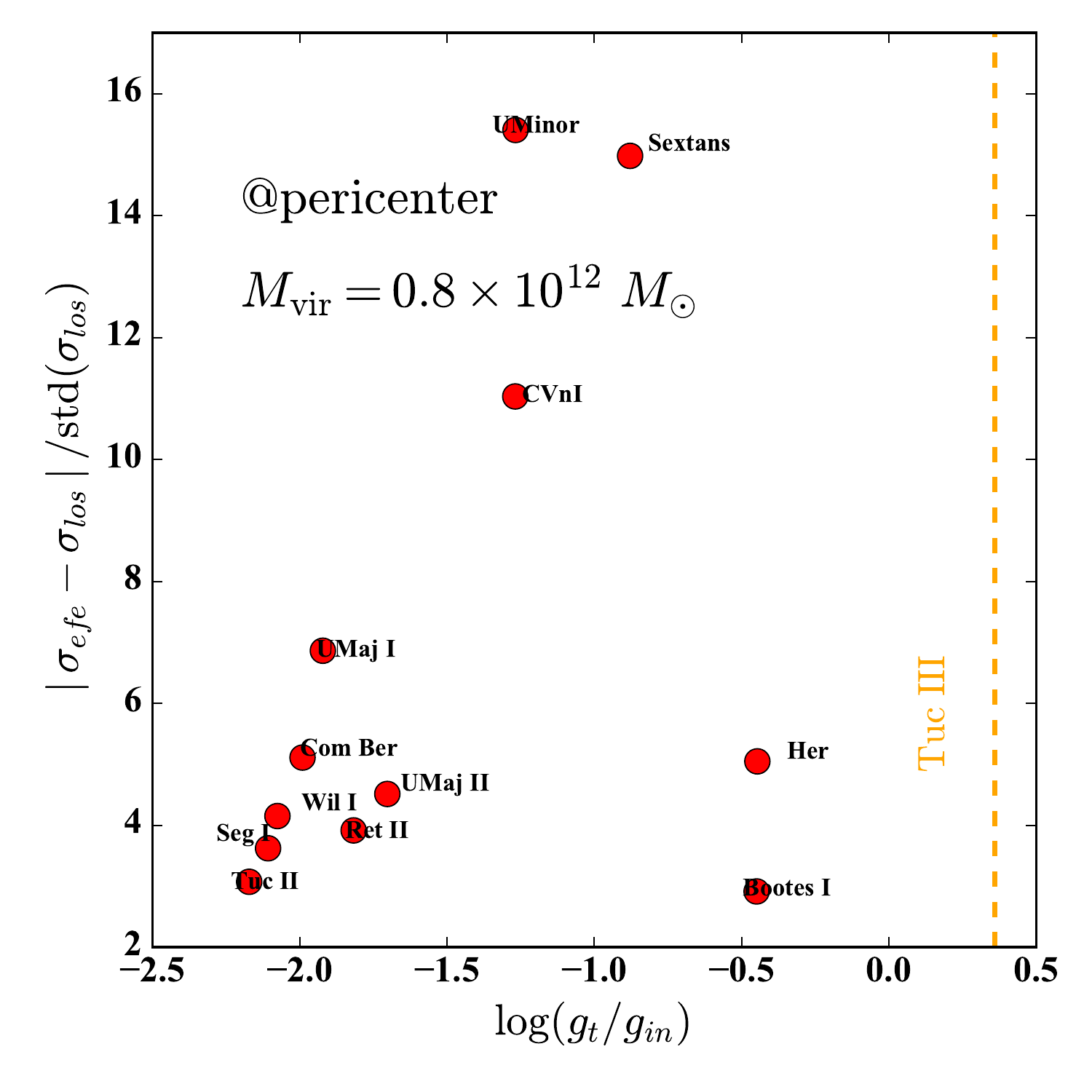}
\includegraphics[width=\columnwidth]{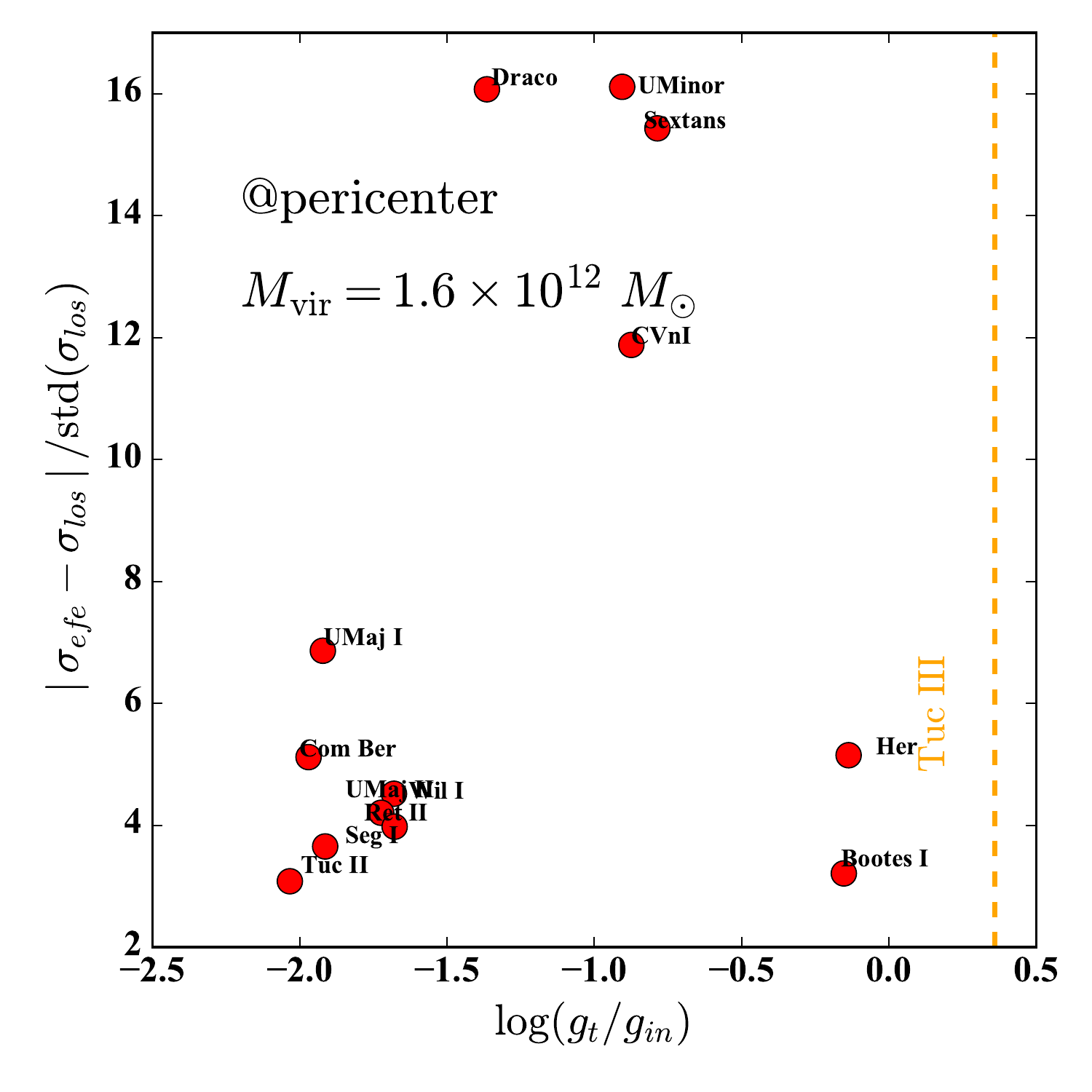}
\caption{Same as in Figure \ref{fig_2} but calculated at the distance of pericenter approach. Top (bottom) panel shows the derived pericenter distance assuming a MW's potential with a virial mass of $0.8\times10^{12}$ ($1.6\times10^{12}$)$\msun$.
The correlation between deviations from MOND's prediction disappears in the limit of low tidal to internal acceleration ratio, which can be a challenging result for MOND to account for. Hercules stands out as a satellite with rather high tidal to internal acceleration ratio at its pericenter passage but not deviating as much as  Draco, U Mi, and Sextans would suggest.}
\label{fig_4}
\end{figure}

So far, all calculations have been done based on the current distance of the satellites from the Galactic center. 
However, we know that the satellites have experienced a pericenter approach, and if their orbit is highly eccentric, tidal acceleration at pericenter passage is important. 
This is shown in Figure \ref{fig_4}. We obtain the estimated pericenter approach from \citet{Fritz2018} assuming two different halo mass for the MW's halo. The obtained estimates from the low virial mass calculation of \citet{Fritz2018} agrees with
the results in \citet{Simon2018ApJGaia}. We note that the agreement is not perfect, and some deviations in the estimated orbital parameters are present. If the existence of a correlation between the MOND's deviation and tidal effects is the desired feature to look for, then the data shows that such a correlation disappears if we focus on satellites with low ratios of tidal to internal acceleration even at pericenter passage. We note that the data point for U Maj I indicates its estimated pericenter approach while the satellite has not reached its pericenter distance. 
Tucana II and Seg I deviate at 3 and 4 $\sigma$ level from MOND's prediction despite having a low tidal to internal acceleration at pericenter approach. 
Coma Berenices stands at 5-$\sigma$ deviation from MOND's prediction while having a rather large pericenter approach of 43 kpcs. However, since this satellite is an early infall system, its deviation could be justified by being in the tidal field of its host for about a Hubble time. Moreover, despite its high ratio of tidal to internal acceleration, Hercules does not deviate from MOND's predictions as much as is expected from Draco, U Min, and Sextans.

\section{Discussion}\label{sec:discussion}

The latest Gaia DR 2 data shows a lack of strong correlation between tidal susceptibility and deviation from MOND's predictions and observed line of sight velocity dispersion of the UFDs, as presented in Figure \ref{fig_2}. 
While the data challenges MOND, more data is needed. In this work, we solely focused on tidal features as a proxy for deviation from MOND. We did not investigate correlations with other estimators such as ellipticity of the UFD as projected on the sky, or adiabaticity (which is a measure of the number of internal orbits of stars over the number of orbits of the satellite around its host).
Ellipticity, adiabaticity, and other estimators have been used in \citet{MW2010ApJ} and shown to be a proxy to tidal features and we do not investigate these in this work.

Could we be biased in inferring the intrinsic velocity dispersion of the UFDs? We know that binary stars can inflate the estimated 1 D velocity dispersion specifically for systems in which the total number of the stars is $\leq 10$. 
For example, a binary star did inflate the velocity dispersion for Triangulum II \citep{Kirby2017ApJ}. However, for well-studied systems such as seg I, binaries contribute to the velocity dispersion at 10\% level \citep{Simon2011ApJSegI}.
In the meantime, a detailed analysis of Draco and U Min shows that the contribution of binaries to the observed velocity dispersion can be significant \citep{Spencer2018AJ}.
While single epoch observations of stars are vulnerable to binaries, future multi-epoch campaigns will say the final word on the issue of binaries. 

It is also worth mentioning the remarkable prediction of MOND for M31 dwarf satellites \citep{MM_M31,MM_M31_2013}. Moreover, the original prediction of MOND for Cetus and Tucana of 8.2 and 5.3 $\pm0.4 \kms$, respectively, was not in agreement with the observations at the time (estimated values to be about $17.2\pm2$ and $15.8^{+4.1}_{-3.1}~\kms$ for Cetus and Tucana, respectively). However, recent observations point to a much better agreement with MOND's prediction with Tucana having a velocity dispersion of $6.2^{+1.6}_{-1.3}~\kms$ \citep{Taibi2020AA}. This can be taken as MOND will survive with time as better data is gathered for the UFDs. 

The issue pertaining to MOND's predictions is not limited to the line of sight velocity dispersion. 
For example, it is known that dynamical friction is stronger in MOND (shorter timescale for a satellite) compared to dark matter models \citep{Niooti2008MNRAS} which makes Fornax and its globular cluster a serious problem for MOND.
However, Fornax is a problem for the standard model of $\rm \Lambda CDM$ too, and a core-like structure for its dark matter halo can allow for the dynamics of globular clusters observed in Fornax. 
We note that other circular orbits outside the tidal radius of Fornax can justify the dynamics of globular clusters in Fornax, even in MOND \citep{Angus2009MNRAS}.  

What do the seemingly simple modification to Newtonian gravity and its success at the galactic scale teach us about the nature of dark matter? Why does MOND have shortcomings at larger scales, such as in clusters of galaxies? And based on the results we analyzed in this work, apparently shortcoming in the scale of UFDs? Is this shortcoming in satellites scale an issue with EFE modeling? 
Attempts to successfully merge these apparently different behaviors of dark matter under one umbrella have been limited \citep[see for example the superfluid interpretation of the dark matter; ][]{Berezhiani2015PhRvD,Khoury2015PhRvD}, 
and the emergent gravity paradigm \citep{2Verlinde017ScPP}, but perhaps such avenues hold the key to understand the nature of dark matter.

We are hopeful that future data obtained by the Vera C. Rubin Observatory, which will find all satellites of the MW within its virial radius down to V band absolute magnitude fainter than $M_V=-4$, 
discover enough UFDs to have a statistically significant sample for testing MOND \citep{Tollerud2008ApJ}.

\acknowledgements 
We are thankful to the referee for their constructive comments. 
We are thankful to Mordehai Milgrom, Stacy McGaugh, and Josh Simon for their insightful comments. This work was supported in part by the Dean’s Competitive Fund for Promising Scholarship at the Faculty of Arts \& Sciences of Harvard University.

\bibliographystyle{yahapj}
\bibliography{ms}

\begin{thebibliography}{}
\providecommand\natexlab[1]{#1}
\providecommand\JournalTitle[1]{#1}

\bibitem[{{Abadi} {et~al.}(2003){Abadi}, {Navarro}, {Steinmetz}, \&
  {Eke}}]{Abadi2003ApJ}
{Abadi}, M.~G., {Navarro}, J.~F., {Steinmetz}, M., \& {Eke}, V.~R. 2003,
  \href{http://dx.doi.org/10.1086/375512}{\JournalTitle{\apj}, 591, 499}

\bibitem[{{Adams} \& {Oosterloo}(2018)}]{Adams2018AA}
{Adams}, E. A.~K., \& {Oosterloo}, T.~A. 2018,
  \href{http://dx.doi.org/10.1051/0004-6361/201732017}{\JournalTitle{\aap},
  612, A26}

\bibitem[{{Angus} \& {Diaferio}(2009)}]{Angus2009MNRAS}
{Angus}, G.~W., \& {Diaferio}, A. 2009,
  \href{http://dx.doi.org/10.1111/j.1365-2966.2009.14745.x}{\JournalTitle{\mnras},
  396, 887}

\bibitem[{{Bechtol} {et~al.}(2015){Bechtol}, {Drlica-Wagner}, {Balbinot},
  {Pieres}, {Simon}, {Yanny}, {Santiago}, {Wechsler}, {Frieman}, {Walker},
  {Williams}, {Rozo}, {Rykoff}, {Queiroz}, {Luque}, {Benoit-L{\'e}vy},
  {Tucker}, {Sevilla}, {Gruendl}, {da Costa}, {Fausti Neto}, {Maia}, {Abbott},
  {Allam}, {Armstrong}, {Bauer}, {Bernstein}, {Bernstein}, {Bertin}, {Brooks},
  {Buckley-Geer}, {Burke}, {Carnero Rosell}, {Castander}, {Covarrubias},
  {D'Andrea}, {DePoy}, {Desai}, {Diehl}, {Eifler}, {Estrada}, {Evrard},
  {Fernandez}, {Finley}, {Flaugher}, {Gaztanaga}, {Gerdes}, {Girardi},
  {Gladders}, {Gruen}, {Gutierrez}, {Hao}, {Honscheid}, {Jain}, {James},
  {Kent}, {Kron}, {Kuehn}, {Kuropatkin}, {Lahav}, {Li}, {Lin}, {Makler},
  {March}, {Marshall}, {Martini}, {Merritt}, {Miller}, {Miquel}, {Mohr},
  {Neilsen}, {Nichol}, {Nord}, {Ogando}, {Peoples}, {Petravick}, {Plazas},
  {Romer}, {Roodman}, {Sako}, {Sanchez}, {Scarpine}, {Schubnell}, {Smith},
  {Soares-Santos}, {Sobreira}, {Suchyta}, {Swanson}, {Tarle}, {Thaler},
  {Thomas}, {Wester}, {Zuntz}, \& {DES Collaboration}}]{Bechtol2015ApJ}
{Bechtol}, K., {Drlica-Wagner}, A., {Balbinot}, E., {et~al.} 2015,
  \href{http://dx.doi.org/10.1088/0004-637X/807/1/50}{\JournalTitle{\apj}, 807,
  50}

\bibitem[{{Berezhiani} \& {Khoury}(2015)}]{Berezhiani2015PhRvD}
{Berezhiani}, L., \& {Khoury}, J. 2015,
  \href{http://dx.doi.org/10.1103/PhysRevD.92.103510}{\JournalTitle{\prd}, 92,
  103510}

\bibitem[{{Bouch{\'e}} {et~al.}(2010){Bouch{\'e}}, {Dekel}, {Genzel}, {Genel},
  {Cresci}, {F{\"o}rster Schreiber}, {Shapiro}, {Davies}, \&
  {Tacconi}}]{Bouch2010ApJ}
{Bouch{\'e}}, N., {Dekel}, A., {Genzel}, R., {et~al.} 2010,
  \href{http://dx.doi.org/10.1088/0004-637X/718/2/1001}{\JournalTitle{\apj},
  718, 1001}

\bibitem[{{Burkert}(2020)}]{Burkert2020ApJ}
{Burkert}, A. 2020,
  \href{http://dx.doi.org/10.3847/1538-4357/abb242}{\JournalTitle{\apj}, 904,
  161}

\bibitem[{{Caldwell} {et~al.}(2017){Caldwell}, {Walker}, {Mateo}, {Olszewski},
  {Koposov}, {Belokurov}, {Torrealba}, {Geringer-Sameth}, \&
  {Johnson}}]{Caldwell2017ApJ}
{Caldwell}, N., {Walker}, M.~G., {Mateo}, M., {et~al.} 2017,
  \href{http://dx.doi.org/10.3847/1538-4357/aa688e}{\JournalTitle{\apj}, 839,
  20}

\bibitem[{{Chan} {et~al.}(2015){Chan}, {Kere{\v{s}}}, {O{\~n}orbe}, {Hopkins},
  {Muratov}, {Faucher-Gigu{\`e}re}, \& {Quataert}}]{Chan2015MNRAS}
{Chan}, T.~K., {Kere{\v{s}}}, D., {O{\~n}orbe}, J., {et~al.} 2015,
  \href{http://dx.doi.org/10.1093/mnras/stv2165}{\JournalTitle{\mnras}, 454,
  2981}

\bibitem[{{Fillingham} {et~al.}(2019){Fillingham}, {Cooper}, {Kelley},
  {Rodriguez Wimberly}, {Boylan-Kolchin}, {Bullock}, {Garrison-Kimmel},
  {Pawlowski}, \& {Wheeler}}]{Fillingham2019}
{Fillingham}, S.~P., {Cooper}, M.~C., {Kelley}, T., {et~al.} 2019,
  \JournalTitle{arXiv e-prints}, arXiv:1906.04180

\bibitem[{{Fritz} {et~al.}(2018){Fritz}, {Battaglia}, {Pawlowski},
  {Kallivayalil}, {van der Marel}, {Sohn}, {Brook}, \& {Besla}}]{Fritz2018}
{Fritz}, T.~K., {Battaglia}, G., {Pawlowski}, M.~S., {et~al.} 2018,
  \href{http://dx.doi.org/10.1051/0004-6361/201833343}{\JournalTitle{\aap},
  619, A103}

\bibitem[{{Gallart} {et~al.}(2021){Gallart}, {Monelli}, {Ruiz-Lara},
  {Calamida}, {Cassisi}, {Cignoni}, {Anderson}, {Battaglia}, {Bermejo-Climent},
  {Bernard}, {Mart{\'\i}nez-V{\'a}zquez}, {Mayer}, {Salvadori}, {Monachesi},
  {Navarro}, {Shen}, {Surot}, {Tosi}, {Bajaj}, \&
  {Strinfellow}}]{Gallart2021ApJ}
{Gallart}, C., {Monelli}, M., {Ruiz-Lara}, T., {et~al.} 2021,
  \href{http://dx.doi.org/10.3847/1538-4357/abddbe}{\JournalTitle{\apj}, 909,
  192}

\bibitem[{{Governato} {et~al.}(2007){Governato}, {Willman}, {Mayer}, {Brooks},
  {Stinson}, {Valenzuela}, {Wadsley}, \& {Quinn}}]{Governato2007MNRAS}
{Governato}, F., {Willman}, B., {Mayer}, L., {et~al.} 2007,
  \href{http://dx.doi.org/10.1111/j.1365-2966.2006.11266.x}{\JournalTitle{\mnras},
  374, 1479}

\bibitem[{{Hayashi} {et~al.}(2021){Hayashi}, {Ferreira}, \& {Jowett
  Chan}}]{Hayashi2021}
{Hayashi}, K., {Ferreira}, E. G.~M., \& {Jowett Chan}, H.~Y. 2021,
  \JournalTitle{arXiv e-prints}, arXiv:2102.05300

\bibitem[{{Jerjen} {et~al.}(2018){Jerjen}, {Conn}, {Kim}, \&
  {Schirmer}}]{Jerjen2018}
{Jerjen}, H., {Conn}, B., {Kim}, D., \& {Schirmer}, M. 2018,
  \JournalTitle{arXiv e-prints}, arXiv:1809.02259

\bibitem[{{Khoury}(2015)}]{Khoury2015PhRvD}
{Khoury}, J. 2015,
  \href{http://dx.doi.org/10.1103/PhysRevD.91.024022}{\JournalTitle{\prd}, 91,
  024022}

\bibitem[{{Kirby} {et~al.}(2013{\natexlab{a}}){Kirby}, {Boylan-Kolchin},
  {Cohen}, {Geha}, {Bullock}, \& {Kaplinghat}}]{Kirby2013ApJSeg2}
{Kirby}, E.~N., {Boylan-Kolchin}, M., {Cohen}, J.~G., {et~al.}
  2013{\natexlab{a}},
  \href{http://dx.doi.org/10.1088/0004-637X/770/1/16}{\JournalTitle{\apj}, 770,
  16}

\bibitem[{{Kirby} {et~al.}(2013{\natexlab{b}}){Kirby}, {Cohen}, {Guhathakurta},
  {Cheng}, {Bullock}, \& {Gallazzi}}]{Kirby2013ApJ}
{Kirby}, E.~N., {Cohen}, J.~G., {Guhathakurta}, P., {et~al.}
  2013{\natexlab{b}},
  \href{http://dx.doi.org/10.1088/0004-637X/779/2/102}{\JournalTitle{\apj},
  779, 102}

\bibitem[{{Kirby} {et~al.}(2017){Kirby}, {Cohen}, {Simon}, {Guhathakurta},
  {Thygesen}, \& {Duggan}}]{Kirby2017ApJ}
{Kirby}, E.~N., {Cohen}, J.~G., {Simon}, J.~D., {et~al.} 2017,
  \href{http://dx.doi.org/10.3847/1538-4357/aa6570}{\JournalTitle{\apj}, 838,
  83}

\bibitem[{{Koposov} {et~al.}(2015{\natexlab{a}}){Koposov}, {Belokurov},
  {Torrealba}, \& {Evans}}]{Koposov2015ApJ}
{Koposov}, S.~E., {Belokurov}, V., {Torrealba}, G., \& {Evans}, N.~W.
  2015{\natexlab{a}},
  \href{http://dx.doi.org/10.1088/0004-637X/805/2/130}{\JournalTitle{\apj},
  805, 130}

\bibitem[{{Koposov} {et~al.}(2011){Koposov}, {Gilmore}, {Walker}, {Belokurov},
  {Evans}, {Fellhauer}, {Gieren}, {Geisler}, {Monaco}, {Norris}, {Okamoto},
  {Pe{\~n}arrubia}, {Wilkinson}, {Wyse}, \& {Zucker}}]{Koposov2011}
{Koposov}, S.~E., {Gilmore}, G., {Walker}, M.~G., {et~al.} 2011,
  \href{http://dx.doi.org/10.1088/0004-637X/736/2/146}{\JournalTitle{\apj},
  736, 146}

\bibitem[{{Koposov} {et~al.}(2015{\natexlab{b}}){Koposov}, {Casey},
  {Belokurov}, {Lewis}, {Gilmore}, {Worley}, {Hourihane}, {Randich}, {Bensby},
  {Bragaglia}, {Bergemann}, {Carraro}, {Costado}, {Flaccomio}, {Francois},
  {Heiter}, {Hill}, {Jofre}, {Lando}, {Lanzafame}, {de Laverny}, {Monaco},
  {Morbidelli}, {Sbordone}, {Mikolaitis}, \& {Ryde}}]{Koposov2015ApJRetIIHolI}
{Koposov}, S.~E., {Casey}, A.~R., {Belokurov}, V., {et~al.} 2015{\natexlab{b}},
  \href{http://dx.doi.org/10.1088/0004-637X/811/1/62}{\JournalTitle{\apj}, 811,
  62}

\bibitem[{{Li} {et~al.}(2017){Li}, {Simon}, {Drlica-Wagner}, {Bechtol}, {Wang},
  {Garc{\'\i}a-Bellido}, {Frieman}, {Marshall}, {James}, {Strigari}, {Pace},
  {Balbinot}, {Zhang}, {Abbott}, {Allam}, {Benoit-L{\'e}vy}, {Bernstein},
  {Bertin}, {Brooks}, {Burke}, {Carnero Rosell}, {Carrasco Kind}, {Carretero},
  {Cunha}, {D'Andrea}, {da Costa}, {DePoy}, {Desai}, {Diehl}, {Eifler},
  {Flaugher}, {Goldstein}, {Gruen}, {Gruendl}, {Gschwend}, {Gutierrez},
  {Krause}, {Kuehn}, {Lin}, {Maia}, {March}, {Menanteau}, {Miquel}, {Plazas},
  {Romer}, {Sanchez}, {Santiago}, {Schubnell}, {Sevilla-Noarbe}, {Smith},
  {Sobreira}, {Suchyta}, {Tarle}, {Thomas}, {Tucker}, {Walker}, {Wechsler},
  {Wester}, {Yanny}, \& {DES Collaboration}}]{Li2017ApJ}
{Li}, T.~S., {Simon}, J.~D., {Drlica-Wagner}, A., {et~al.} 2017,
  \href{http://dx.doi.org/10.3847/1538-4357/aa6113}{\JournalTitle{\apj}, 838,
  8}

\bibitem[{{Li} {et~al.}(2018){Li}, {Simon}, {Kuehn}, {Pace}, {Erkal},
  {Bechtol}, {Yanny}, {Drlica-Wagner}, {Marshall}, {Lidman}, {Balbinot},
  {Carollo}, {Jenkins}, {Mart{\'\i}nez-V{\'a}zquez}, {Shipp}, {Stringer},
  {Vivas}, {Walker}, {Wechsler}, {Abdalla}, {Allam}, {Annis}, {Avila},
  {Bertin}, {Brooks}, {Buckley-Geer}, {Burke}, {Carnero Rosell}, {Carrasco
  Kind}, {Carretero}, {Cunha}, {D'Andrea}, {da Costa}, {Davis}, {De Vicente},
  {Doel}, {Eifler}, {Evrard}, {Flaugher}, {Frieman}, {Garc{\'\i}a-Bellido},
  {Gaztanaga}, {Gerdes}, {Gruen}, {Gruendl}, {Gschwend}, {Gutierrez},
  {Hartley}, {Hollowood}, {Honscheid}, {James}, {Krause}, {Maia}, {March},
  {Menanteau}, {Miquel}, {Plazas}, {Sanchez}, {Santiago}, {Scarpine},
  {Schindler}, {Schubnell}, {Sevilla-Noarbe}, {Smith}, {Smith},
  {Soares-Santos}, {Sobreira}, {Suchyta}, {Swanson}, {Tarle}, {Tucker}, \& {DES
  Collaboration}}]{Li2018ApJ}
{Li}, T.~S., {Simon}, J.~D., {Kuehn}, K., {et~al.} 2018,
  \href{http://dx.doi.org/10.3847/1538-4357/aadf91}{\JournalTitle{\apj}, 866,
  22}

\bibitem[{{McConnachie}(2012)}]{McConnachie2012}
{McConnachie}, A.~W. 2012,
  \href{http://dx.doi.org/10.1088/0004-6256/144/1/4}{\JournalTitle{\aj}, 144,
  4}

\bibitem[{{McGaugh} \& {Milgrom}(2013{\natexlab{a}})}]{MM2013ApJ}
{McGaugh}, S., \& {Milgrom}, M. 2013{\natexlab{a}},
  \href{http://dx.doi.org/10.1088/0004-637X/766/1/22}{\JournalTitle{\apj}, 766,
  22}

\bibitem[{{McGaugh} \& {Milgrom}(2013{\natexlab{b}})}]{MM_M31}
---. 2013{\natexlab{b}},
  \href{http://dx.doi.org/10.1088/0004-637X/766/1/22}{\JournalTitle{\apj}, 766,
  22}

\bibitem[{{McGaugh} \& {Milgrom}(2013{\natexlab{c}})}]{MM_M31_2013}
---. 2013{\natexlab{c}},
  \href{http://dx.doi.org/10.1088/0004-637X/775/2/139}{\JournalTitle{\apj},
  775, 139}

\bibitem[{{McGaugh}(2005)}]{McGaugh2005ApJ}
{McGaugh}, S.~S. 2005,
  \href{http://dx.doi.org/10.1086/432968}{\JournalTitle{\apj}, 632, 859}

\bibitem[{{McGaugh}(2016)}]{McGaugh2016ApJ}
---. 2016,
  \href{http://dx.doi.org/10.3847/2041-8205/832/1/L8}{\JournalTitle{\apjl},
  832, L8}

\bibitem[{{McGaugh} {et~al.}(2000){McGaugh}, {Schombert}, {Bothun}, \& {de
  Blok}}]{McGaugh2000ApJ}
{McGaugh}, S.~S., {Schombert}, J.~M., {Bothun}, G.~D., \& {de Blok}, W.~J.~G.
  2000, \href{http://dx.doi.org/10.1086/312628}{\JournalTitle{\apjl}, 533, L99}

\bibitem[{{McGaugh} \& {Wolf}(2010)}]{MW2010ApJ}
{McGaugh}, S.~S., \& {Wolf}, J. 2010,
  \href{http://dx.doi.org/10.1088/0004-637X/722/1/248}{\JournalTitle{\apj},
  722, 248}

\bibitem[{{Milgrom}(1983)}]{Milgrom1983ApJ}
{Milgrom}, M. 1983,
  \href{http://dx.doi.org/10.1086/161130}{\JournalTitle{\apj}, 270, 365}

\bibitem[{{Mutlu-Pakdil} {et~al.}(2018){Mutlu-Pakdil}, {Sand}, {Carlin},
  {Spekkens}, {Caldwell}, {Crnojevi{\'c}}, {Hughes}, {Willman}, \&
  {Zaritsky}}]{MutluPakdil2018ApJ}
{Mutlu-Pakdil}, B., {Sand}, D.~J., {Carlin}, J.~L., {et~al.} 2018,
  \href{http://dx.doi.org/10.3847/1538-4357/aacd0e}{\JournalTitle{\apj}, 863,
  25}

\bibitem[{{Nadler} {et~al.}(2021){Nadler}, {Drlica-Wagner}, {Bechtol}, {Mau},
  {Wechsler}, {Gluscevic}, {Boddy}, {Pace}, {Li}, {McNanna}, {Riley},
  {Garc{\'\i}a-Bellido}, {Mao}, {Green}, {Burke}, {Peter}, {Jain}, {Abbott},
  {Aguena}, {Allam}, {Annis}, {Avila}, {Brooks}, {Carrasco Kind}, {Carretero},
  {Costanzi}, {da Costa}, {De Vicente}, {Desai}, {Diehl}, {Doel}, {Everett},
  {Evrard}, {Flaugher}, {Frieman}, {Gerdes}, {Gruen}, {Gruendl}, {Gschwend},
  {Gutierrez}, {Hinton}, {Honscheid}, {Huterer}, {James}, {Krause}, {Kuehn},
  {Kuropatkin}, {Lahav}, {Maia}, {Marshall}, {Menanteau}, {Miquel}, {Palmese},
  {Paz-Chinch{\'o}n}, {Plazas}, {Romer}, {Sanchez}, {Scarpine}, {Serrano},
  {Sevilla-Noarbe}, {Smith}, {Soares-Santos}, {Suchyta}, {Swanson}, {Tarle},
  {Tucker}, {Walker}, {Wester}, \& {DES Collaboration}}]{Nadler2021PhRvL}
{Nadler}, E.~O., {Drlica-Wagner}, A., {Bechtol}, K., {et~al.} 2021,
  \href{http://dx.doi.org/10.1103/PhysRevLett.126.091101}{\JournalTitle{\prl},
  126, 091101}

\bibitem[{{Nagasawa} {et~al.}(2018){Nagasawa}, {Marshall}, {Li}, {Hansen},
  {Simon}, {Bernstein}, {Balbinot}, {Drlica-Wagner}, {Pace}, {Strigari},
  {Pellegrino}, {DePoy}, {Suntzeff}, {Bechtol}, {Walker}, {Abbott}, {Abdalla},
  {Allam}, {Annis}, {Benoit-L{\'e}vy}, {Bertin}, {Brooks}, {Carnero Rosell},
  {Carrasco Kind}, {Carretero}, {Cunha}, {D'Andrea}, {da Costa}, {Davis},
  {Desai}, {Doel}, {Eifler}, {Flaugher}, {Fosalba}, {Frieman},
  {Garc{\'\i}a-Bellido}, {Gaztanaga}, {Gerdes}, {Gruen}, {Gruendl}, {Gschwend},
  {Gutierrez}, {Hartley}, {Honscheid}, {James}, {Jeltema}, {Krause}, {Kuehn},
  {Kuhlmann}, {Kuropatkin}, {March}, {Miquel}, {Nord}, {Roodman}, {Sanchez},
  {Santiago}, {Scarpine}, {Schindler}, {Schubnell}, {Sevilla-Noarbe}, {Smith},
  {Smith}, {Soares-Santos}, {Sobreira}, {Suchyta}, {Tarle}, {Thomas}, {Tucker},
  {Wechsler}, {Wolf}, \& {Yanny}}]{Nagasawa2018ApJ}
{Nagasawa}, D.~Q., {Marshall}, J.~L., {Li}, T.~S., {et~al.} 2018,
  \href{http://dx.doi.org/10.3847/1538-4357/aaa01d}{\JournalTitle{\apj}, 852,
  99}

\bibitem[{{Nipoti} {et~al.}(2008){Nipoti}, {Ciotti}, {Binney}, \&
  {Londrillo}}]{Niooti2008MNRAS}
{Nipoti}, C., {Ciotti}, L., {Binney}, J., \& {Londrillo}, P. 2008,
  \href{http://dx.doi.org/10.1111/j.1365-2966.2008.13192.x}{\JournalTitle{\mnras},
  386, 2194}

\bibitem[{{Profumo} {et~al.}(2019){Profumo}, {Giani}, \&
  {Piattella}}]{Profumo2019}
{Profumo}, S., {Giani}, L., \& {Piattella}, O.~F. 2019,
  \href{http://dx.doi.org/10.3390/universe5100213}{\JournalTitle{Universe}, 5,
  213}

\bibitem[{{Reid} {et~al.}(2014){Reid}, {Menten}, {Brunthaler}, {Zheng}, {Dame},
  {Xu}, {Wu}, {Zhang}, {Sanna}, {Sato}, {Hachisuka}, {Choi}, {Immer},
  {Moscadelli}, {Rygl}, \& {Bartkiewicz}}]{Reid2014ApJ}
{Reid}, M.~J., {Menten}, K.~M., {Brunthaler}, A., {et~al.} 2014,
  \href{http://dx.doi.org/10.1088/0004-637X/783/2/130}{\JournalTitle{\apj},
  783, 130}

\bibitem[{{Rocha} {et~al.}(2012){Rocha}, {Peter}, \&
  {Bullock}}]{Rocha2012MNRAS}
{Rocha}, M., {Peter}, A. H.~G., \& {Bullock}, J. 2012,
  \href{http://dx.doi.org/10.1111/j.1365-2966.2012.21432.x}{\JournalTitle{\mnras},
  425, 231}

\bibitem[{{Rubin} {et~al.}(1980){Rubin}, {Ford}, \& {Thonnard}}]{Rubin1980ApJ}
{Rubin}, V.~C., {Ford}, W.~K., J., \& {Thonnard}, N. 1980,
  \href{http://dx.doi.org/10.1086/158003}{\JournalTitle{\apj}, 238, 471}

\bibitem[{{Safarzadeh} \& {Spergel}(2020)}]{SS2020ApJ}
{Safarzadeh}, M., \& {Spergel}, D.~N. 2020,
  \href{http://dx.doi.org/10.3847/1538-4357/ab7db2}{\JournalTitle{\apj}, 893,
  21}

\bibitem[{{Sales} {et~al.}(2017){Sales}, {Navarro}, {Oman}, {Fattahi},
  {Ferrero}, {Abadi}, {Bower}, {Crain}, {Frenk}, {Sawala}, {Schaller},
  {Schaye}, {Theuns}, \& {White}}]{Sales2017MNRAS}
{Sales}, L.~V., {Navarro}, J.~F., {Oman}, K., {et~al.} 2017,
  \href{http://dx.doi.org/10.1093/mnras/stw2461}{\JournalTitle{\mnras}, 464,
  2419}

\bibitem[{{Sanders} \& {McGaugh}(2002)}]{SM02}
{Sanders}, R.~H., \& {McGaugh}, S.~S. 2002,
  \href{http://dx.doi.org/10.1146/annurev.astro.40.060401.093923}{\JournalTitle{\araa},
  40, 263}

\bibitem[{{Simon}(2018)}]{Simon2018ApJGaia}
{Simon}, J.~D. 2018,
  \href{http://dx.doi.org/10.3847/1538-4357/aacdfb}{\JournalTitle{\apj}, 863,
  89}

\bibitem[{{Simon}(2019)}]{Simon2019ARAA}
---. 2019,
  \href{http://dx.doi.org/10.1146/annurev-astro-091918-104453}{\JournalTitle{\araa},
  57, 375}

\bibitem[{{Simon} \& {Geha}(2007)}]{SG2007ApJ}
{Simon}, J.~D., \& {Geha}, M. 2007,
  \href{http://dx.doi.org/10.1086/521816}{\JournalTitle{\apj}, 670, 313}

\bibitem[{{Simon} {et~al.}(2011){Simon}, {Geha}, {Minor}, {Martinez}, {Kirby},
  {Bullock}, {Kaplinghat}, {Strigari}, {Willman}, {Choi}, {Tollerud}, \&
  {Wolf}}]{Simon2011ApJSegI}
{Simon}, J.~D., {Geha}, M., {Minor}, Q.~E., {et~al.} 2011,
  \href{http://dx.doi.org/10.1088/0004-637X/733/1/46}{\JournalTitle{\apj}, 733,
  46}

\bibitem[{{Simon} {et~al.}(2015){Simon}, {Drlica-Wagner}, {Li}, {Nord}, {Geha},
  {Bechtol}, {Balbinot}, {Buckley-Geer}, {Lin}, {Marshall}, {Santiago},
  {Strigari}, {Wang}, {Wechsler}, {Yanny}, {Abbott}, {Bauer}, {Bernstein},
  {Bertin}, {Brooks}, {Burke}, {Capozzi}, {Carnero Rosell}, {Carrasco Kind},
  {D'Andrea}, {da Costa}, {DePoy}, {Desai}, {Diehl}, {Dodelson}, {Cunha},
  {Estrada}, {Evrard}, {Fausti Neto}, {Fernandez}, {Finley}, {Flaugher},
  {Frieman}, {Gaztanaga}, {Gerdes}, {Gruen}, {Gruendl}, {Honscheid}, {James},
  {Kent}, {Kuehn}, {Kuropatkin}, {Lahav}, {Maia}, {March}, {Martini}, {Miller},
  {Miquel}, {Ogando}, {Romer}, {Roodman}, {Rykoff}, {Sako}, {Sanchez},
  {Schubnell}, {Sevilla}, {Smith}, {Soares-Santos}, {Sobreira}, {Suchyta},
  {Swanson}, {Tarle}, {Thaler}, {Tucker}, {Vikram}, {Walker}, {Wester}, \& {DES
  Collaboration}}]{Simon2015ApJRetII}
{Simon}, J.~D., {Drlica-Wagner}, A., {Li}, T.~S., {et~al.} 2015,
  \href{http://dx.doi.org/10.1088/0004-637X/808/1/95}{\JournalTitle{\apj}, 808,
  95}

\bibitem[{{Simon} {et~al.}(2017){Simon}, {Li}, {Drlica-Wagner}, {Bechtol},
  {Marshall}, {James}, {Wang}, {Strigari}, {Balbinot}, {Kuehn}, {Walker},
  {Abbott}, {Allam}, {Annis}, {Benoit-L{\'e}vy}, {Brooks}, {Buckley-Geer},
  {Burke}, {Carnero Rosell}, {Carrasco Kind}, {Carretero}, {Cunha}, {D'Andrea},
  {da Costa}, {DePoy}, {Desai}, {Doel}, {Fernandez}, {Flaugher}, {Frieman},
  {Garc{\'\i}a-Bellido}, {Gaztanaga}, {Goldstein}, {Gruen}, {Gutierrez},
  {Kuropatkin}, {Maia}, {Martini}, {Menanteau}, {Miller}, {Miquel}, {Neilsen},
  {Nord}, {Ogando}, {Plazas}, {Romer}, {Rykoff}, {Sanchez}, {Santiago},
  {Scarpine}, {Schubnell}, {Sevilla-Noarbe}, {Smith}, {Sobreira}, {Suchyta},
  {Swanson}, {Tarle}, {Whiteway}, {Yanny}, \& {DES
  Collaboration}}]{Simon2017ApJTucIII}
{Simon}, J.~D., {Li}, T.~S., {Drlica-Wagner}, A., {et~al.} 2017,
  \href{http://dx.doi.org/10.3847/1538-4357/aa5be7}{\JournalTitle{\apj}, 838,
  11}

\bibitem[{{Simon} {et~al.}(2020){Simon}, {Li}, {Erkal}, {Pace},
  {Drlica-Wagner}, {James}, {Marshall}, {Bechtol}, {Hansen}, {Kuehn}, {Lidman},
  {Allam}, {Annis}, {Avila}, {Bertin}, {Brooks}, {Burke}, {Rosell}, {Carrasco
  Kind}, {Carretero}, {da Costa}, {De Vicente}, {Desai}, {Doel}, {Eifler},
  {Everett}, {Fosalba}, {Frieman}, {Garc{\'\i}a-Bellido}, {Gaztanaga},
  {Gerdes}, {Gruen}, {Gruendl}, {Gschwend}, {Gutierrez}, {Hollowood},
  {Honscheid}, {Krause}, {Kuropatkin}, {MacCrann}, {Maia}, {March}, {Miquel},
  {Palmese}, {Paz-Chinch{\'o}n}, {Plazas}, {Reil}, {Roodman}, {Sanchez},
  {Santiago}, {Scarpine}, {Schubnell}, {Serrano}, {Smith}, {Suchyta}, {Tarle},
  {Walker}, \& {DES Collaboration}}]{Simon2020ApJ}
{Simon}, J.~D., {Li}, T.~S., {Erkal}, D., {et~al.} 2020,
  \href{http://dx.doi.org/10.3847/1538-4357/ab7ccb}{\JournalTitle{\apj}, 892,
  137}

\bibitem[{{Spencer} {et~al.}(2018){Spencer}, {Mateo}, {Olszewski}, {Walker},
  {McConnachie}, \& {Kirby}}]{Spencer2018AJ}
{Spencer}, M.~E., {Mateo}, M., {Olszewski}, E.~W., {et~al.} 2018,
  \href{http://dx.doi.org/10.3847/1538-3881/aae3e4}{\JournalTitle{\aj}, 156,
  257}

\bibitem[{{Taibi} {et~al.}(2020){Taibi}, {Battaglia}, {Rejkuba}, {Leaman},
  {Kacharov}, {Iorio}, {Jablonka}, \& {Zoccali}}]{Taibi2020AA}
{Taibi}, S., {Battaglia}, G., {Rejkuba}, M., {et~al.} 2020,
  \href{http://dx.doi.org/10.1051/0004-6361/201937240}{\JournalTitle{\aap},
  635, A152}

\bibitem[{{Tollerud} {et~al.}(2008){Tollerud}, {Bullock}, {Strigari}, \&
  {Willman}}]{Tollerud2008ApJ}
{Tollerud}, E.~J., {Bullock}, J.~S., {Strigari}, L.~E., \& {Willman}, B. 2008,
  \href{http://dx.doi.org/10.1086/592102}{\JournalTitle{\apj}, 688, 277}

\bibitem[{{Tully} \& {Fisher}(1977)}]{BTFR}
{Tully}, R.~B., \& {Fisher}, J.~R. 1977, \JournalTitle{\aap}, 500, 105

\bibitem[{{Verlinde}(2017)}]{2Verlinde017ScPP}
{Verlinde}, E. 2017,
  \href{http://dx.doi.org/10.21468/SciPostPhys.2.3.016}{\JournalTitle{SciPost
  Physics}, 2, 016}

\bibitem[{{Vogelsberger} {et~al.}(2013){Vogelsberger}, {Genel}, {Sijacki},
  {Torrey}, {Springel}, \& {Hernquist}}]{Vogelsberger2013MNRAS}
{Vogelsberger}, M., {Genel}, S., {Sijacki}, D., {et~al.} 2013,
  \href{http://dx.doi.org/10.1093/mnras/stt1789}{\JournalTitle{\mnras}, 436,
  3031}

\bibitem[{{Wolf} {et~al.}(2010){Wolf}, {Martinez}, {Bullock}, {Kaplinghat},
  {Geha}, {Mu{\~n}oz}, {Simon}, \& {Avedo}}]{Wolf2010}
{Wolf}, J., {Martinez}, G.~D., {Bullock}, J.~S., {et~al.} 2010,
  \href{http://dx.doi.org/10.1111/j.1365-2966.2010.16753.x}{\JournalTitle{\mnras},
  406, 1220}

\end{thebibliography}
\end{document}